\documentclass[aip,jmp,nofootinbib]{revtex4-1}
\usepackage{dcolumn}
\usepackage{bm}
\usepackage{mathrsfs}
\usepackage{pstricks}
\usepackage{color}
\usepackage{slashed}
\usepackage{amsmath}
\usepackage{epsfig}
\usepackage{amsfonts}
\usepackage{amssymb}
\def\beq{\begin{equation}}
\def\eeq{\end{equation}}
\def\bea{\begin{eqnarray}}
\def\eea{\end{eqnarray}}

\def\Li{\textrm{Li}}
\def\Re{\textrm{Re}}

\def\diag{\textrm{diag}}

\begin{document}

\title{Quantum Field Theories and Prime Numbers Spectrum}
\author{G. Menezes}
\email{gsm@ift.unesp.br}
\affiliation{Instituto de F\'{i}sica Te\'orica, Universidade Estadual Paulista, S\~ao Paulo, SP 01140-070, Brazil}
\author{N. F. Svaiter}
\email{nfuxsvai@cbpf.br}
\affiliation{Centro Brasileiro de Pesquisas F\'{\i}sicas, Rio de Janeiro, RJ 22290-180,
Brazil}

\begin{abstract}

The Riemann hypothesis states that all nontrivial zeros of the zeta function lie on the critical line $\Re(s)=1/2$. Hilbert and P\'olya suggested a possible approach to prove it, based on spectral theory. Within this context, some authors formulated the question: is there a quantum mechanical system related to the sequence of prime numbers? In this Letter we show that such a sequence is not zeta regularizable. Therefore, there are no physical systems described by self-adjoint operators with countably infinite number of degrees of freedom with spectra given by the sequence of primes numbers.
\end{abstract}



\maketitle

\section{Introduction}

Prime numbers occur in a very irregular way within
the sequences of natural numbers.  In particular, the distribution of prime numbers
exhibits a local irregularity but a global regularity.
The best result that we have
concerning their global distribution is the prime number theorem:
the number of primes $\pi(x)$  not exceeding $x$ is asymptotic to $x/\ln x$~\cite{hadamard,c1,c2,ne,kore,ingram}.

The Riemann hypothesis states that all nontrivial zeros of the Riemann zeta function  $\zeta(s)$ lie on the critical line $\Re(s)=1/2$~\cite{riem}. Since in the region of the complex plane where it converges absolutely and uniformly there is a representation for $\zeta(s)$ in terms of the product of all prime numbers, there is a connection between these nontrivial zeros
of the $\zeta(s)$ function and the distribution of prime numbers in the sequence of integers. One way to connect the distribution of primes and the non-trivial zeros of the Riemann zeta function is given by the integral logarithm function $\Li(x)$. The assertion that $\pi(x)=\Li(x)+O(\sqrt{x}\ln x)$ for $x\rightarrow\infty$ is equivalent to the Riemann hypothesis~\cite{titchmarsh}.

Hilbert and P\'olya suggested that there might be a spectral interpretation for the non-trivial zeros of the Riemann
zeta function. The nontrivial zeros could be the eigenvalues of a self-adjoint linear operator in an appropriate Hilbert space.
The existence of such operator leads us to the proof of the Riemann hypothesis. The spectral nature of the nontrivial zeros emerges studying the local statistical fluctuations of high zeros. These fluctuations are universal and follow the law of fluctuation of the eigenvalue distribution for matrices in the Gaussian unitary ensemble~\cite{mehta}, a result that prompted many authors to consider the Riemann hypothesis in the light of random matrix theory.

Since the main point of Riemann's original paper is that the two sequences, of prime numbers on one hand, and of Riemann zeros on the other hand, are in duality, any statement about one of these two sequences can have a translation in terms of the other. Therefore some authors formulated the following question: is there a quantum mechanical system related to the prime numbers~\cite{mussardo,rosu,zyl,pre}? The idea is to assume that there is a trace formula for the prime numbers and therefore a Hamiltonian with a prime numbers spectrum. For a nice discussion concerning physics of the Riemann hypothesis, see Ref.~\cite{rev}.

Quantum field theory is the formalism where the probabilistic interpretation of quantum mechanics and the special theory of relativity were gathered to account for a plethora of phenomena that cannot be described by quantum mechanics of systems with finitely many degrees of freedom. Euclidean methods lead to mathematically well-defined functional integral representation of the solutions of field theory models. Spectral properties of a Hamiltonian and momentum operators of a field theory can be deduced from cluster properties of moment of the measure that determines the functional integral. Furthermore, boson field theory in one dimension (time) reduces to quantum mechanics formulated in the Heisenberg picture. That is, any quantum mechanical system with $n$ degrees of freedom derives from some $n$-component scalar field model in one dimension. A stochastic process canonically associated with a Hamiltonian $H_{0}+V$ corresponds to Euclidean field self-interaction via the potential $V$~\cite{gert}.

Based on above discussions, one might be tempted to inquire whether quantum field theory techniques can be used to exploit questions related to number theory. In such a direction, connections between field theory and prime numbers have already been investigated by many authors~\cite{sss,bakas,julia,spector}. The aim of this Letter is to investigate the possibility of existence of a quantum field possessing a spectrum given by the sequence of prime numbers. For simplicity, we consider a free scalar field in Euclidean space, but the results derived here can be extended to higher spin fields. We assume that there is a hypothetical class of physical systems described by self-adjoint operators associated with bosonic fields with a spectrum given by the sequence of primes numbers. We show that the functional integrals associated with such theories are ill-defined objects. Here we use $\hbar=c=1$.

\section{Field theory with prime numbers spectrum}

The Riemann zeta function $\zeta(s)$ is a complex variable function of $s$, i.e., $s=\sigma+i\tau$ with $\sigma,\tau \in \mathbb{R}$,
which is defined as the analytic continuation of a Dirichlet series to the whole complex plane. Its unique singularity is the point $s=1$ at which it has a simple pole with unit residue. In the half-plane $\Re(s)>1$ it is defined by the absolutely convergent series
\begin{equation}
\zeta(s)=\sum_{n=1}^{\infty}\,\frac{1}{n^{s}}.
\label{p2}
\end{equation}
In the complex plane for $\Re(s)>1$ the product of all prime numbers appears as a representation of the
Riemann zeta function:
\begin{equation}
\zeta(s)=\prod_{p}\,\Biggl(\frac{1}{1-p^{-s}}\Biggr).
\label{p4}
\end{equation}
The product given by Eq.~(\ref{p4}) is called the Euler product. This is an analytic form of the
fundamental theorem of arithmetic. Since a convergent infinite product never vanishes, from Eq.~(\ref{p4}) we obtain that $\zeta(s)$ has no zeros for $\Re(s)>1$. The proof that the zero-free region includes $\Re(s)=1$,
i.e., $\zeta(1+i\tau)\neq 0$
is another way to formulate the prime number theorem.
From the functional equation
\begin{equation}
\pi^{-\frac{s}{2}}\Gamma\biggl(\frac{s}{2}\biggr)\,\zeta(s)=\pi^{-\frac{(1-s)}{2}}
\Gamma\biggl(\frac{1-s}{2}\biggr)\,\zeta(1-s),
\end{equation}
valid for $s\in \mathbb{C} \setminus\left\{0,1\right\}$ it is possible to show
that the region $\Re(s)<0$ is also zero-free with
respect to the non-trivial zeros.
It is often more convenient to write it as
\begin{equation}
\zeta(s)=\frac{\pi^{\frac{s}{2}}}{\Gamma(\frac{s}{2}+1)}
\frac{1}{(s-1)}\,\xi(s). \label{p6}
\end{equation}
The function $\xi(s)$ is analytic everywhere on the complex plane
and the poles of the Gamma function reproduce all the zeros of
$\zeta(s)$ on the real axis at $\sigma=-2k,\,\,\,k=1,2,...$. Thus
$\xi(s)$ contains only the complex zeroes of $\zeta(s)$.
There is no available proof that there are no other nontrivial complex zeros away
from $\sigma=1/2$.

Now let us construct a field theory satisfying the condition of a prime numbers spectrum. Consider a free neutral scalar field defined in a $d$-dimensional Minkowski spacetime. The Euclidean field theory can be obtained by analytic
continuation of the $n$-point Wightman functions to imaginary time. Let us assume a compact Euclidean space with or without boundary.  If the compact Euclidean space has a smooth boundary, one imposes suitable boundary conditions on the fields.
Let us suppose that there exists an elliptic, self-adjoint differential operator $D$ acting on scalar functions on the
Euclidean space. The usual example is $D=(-\Delta+m_{0}^{2}\,)$, where $\Delta$ is the $d$-dimensional Laplacian.~The kernel $K(m_{0};\,x-y)$ is defined by
\begin{equation}
K(m_{0};x-y)=\left(-\Delta+m_{0}^{2}\,\right)\delta^{d}(x-y).
\label{kernel2}
\end{equation}
The Euclidean generating functional $Z[h]$ is formally defined by the following functional integral:
\begin{equation}
Z[h]=\int [d\varphi]\,\, \exp\left(-S_{0}+ \int d^{d}x\,
h(x)\varphi(x)\right),
\label{1}
\end{equation}
where the action that usually describes a free scalar field is
\begin{equation}
S_{0}(\varphi)=\int d^{d}x\,d^{d}y\,\varphi(x)K(m_{0};\,x-y)\varphi(y).
\label{2}
\end{equation}
In Eq.~(\ref{1}), $[d\varphi]$ is a translational invariant measure, formally given by
$[d\varphi]=\prod_{x} d\varphi(x)$. The term $m_{0}^{2}$ is the (bare) mass squared of the model.
Finally, $h(x)$ is a smooth function introduced to generate the Schwinger functions of the theory.
One can define the functional $W[h]=\ln Z[h]$ which generates the connected Schwinger functions.
To obtain a well-defined object, we need to regularize a determinant associated with the operator $D$, since $W[0]= - 1/2 \ln\det D$. A similar situation arises for the case of self-interacting scalar fields when one calculates the one-loop effective action. For both cases, a finite result can be achieved by defining the spectral zeta function associated with the corresponding elliptic operator~\cite{se,ha,vo}.

The standard technique of the spectral theory of elliptic operators implies that there is a complete orthonormal set $\left\{f_{k}\right\}_{k=1}^{\infty}$ such that the sequence of eigenvalues obey: $0 \leq\lambda_{1}\leq\lambda_{2}\leq
\,...\,\leq\lambda_{k}\rightarrow\infty$, when $k\rightarrow\infty$, where the zero eigenvalue must be omitted (eigenvalues being counted with their multiplicities). In the basis $\left\{f_{k}\right\}$ the operator $D$ is represented by an infinite diagonal matrix $D=\diag\,(\lambda_{1}, \lambda_{2},...)$. Therefore the generic operator
$D$ satisfies $D f_{n}(x)=\lambda_{n}f_{n}(x)$. The spectral zeta function associated with the operator $D$ is defined as
\begin{equation}
\zeta_{D}(s)=\sum_{n}\frac{1}{\lambda_{n}^{s}},\,\,\,\,\,\,\,\Re(s)>s_{0},
\label{imp}
\end{equation}
for some $s_0$. Formally we have
\begin{equation}
-\frac{d}{ds}\zeta_{D}(s)|_{s=0}=\ln\det D.
\label{imp2}
\end{equation}
From equation~(\ref{imp2}), one notes that the spectral zeta function has to be analytically continued beyond the domain of convergence of the series. In other words, it is necessary to perform an analytic continuation
of $\zeta_{D}(s)$ from an open connected set of points for sufficient large positive $\Re(s)$ into the whole
complex plane. In particular, the spectral zeta function must be analytic in a complex neighborhood of $s=0$. This method can also be employed in the case of interacting fields in curved space, at least in a ultrastatic space-time. In this case the one-loop effective action is computed by evaluating a determinant, as mentioned above. Since the argument of the logarithm must be strictly dimensionless, one may have scaling properties:
\begin{equation}
\frac{1}{2}\frac{d}{ds}\zeta_{\mu^{-2}D}(s)|_{s=0}=
\frac{1}{2}\ln\,\mu^{2}
\zeta_{D}|_{s=0}+\frac{1}{2}\frac{d}{ds}\zeta_{D}|_{s=0},
\label{17}
\end{equation}
where $\mu$ is an arbitrary parameter with dimensions of mass.

Let us assume that there is a  hypothetical operator $D$ acting on scalar functions in the Euclidean manifold
with prime numbers as its spectrum. We are using the idea of relating a numerical sequence with the
spectrum of a differential operator. In this case, the spectral zeta function associated with the hypothetical operator $D$
is the so-called prime zeta function $P(s)$, $s=\sigma+i\tau $, $\sigma,\tau \in \mathbb{R}$, defined as
\begin{equation}
P(s)=\sum_{\left\{p\right\}}\,p^{-s},  \,\,\Re(s)>1.
\label{21}
\end{equation}
The above summation is performed over the sequence of all primes~\cite{lan,carl}. Since $\ln(1-1/p)$ is of order $-1/p$ and $\prod_{p}\,(1-p^{-1})=0$, the sum of the reciprocals of the prime numbers diverges, but the series converges absolutely when $\sigma>1$. From the above discussions, in order to regularize the generating functional of connected Schwinger functions, we must analytically extend $P(s)$ to $s=0$. Using the generalization for
the Euler result given by Eq. (\ref{p4}) and the definition of the prime zeta function we have
\begin{equation}
\ln \zeta(s)=\sum_{r=1}^{\infty}\frac{1}{r}\,P(rs),  \,\,\,\Re(s)>1.
\label{23}
\end{equation}
Employing the M\"obius inversion formula, one has
\begin{equation}
P(s)=\sum_{k=1}^{\infty}\frac{\mu(k)}{k}\,\ln\zeta(ks), \,\,\,\Re(s)>1,
\label{24}
\end{equation}
where the M\"obius function $\mu(n)$ is defined as~\cite{hardy}
\begin{displaymath}
\mu(n)=\left\{
\begin{array}{ll}
1\,\,& \mbox{if n is a square-free positive integer}\\
&\mbox{with an even number of prime factors,}\\
-1\,\,& \mbox{if n is a square-free positive integer}\\
&\mbox{with an odd number of prime factors,}\\
0\,\,& \mbox{if n is not square-free,}\\
\end{array}\right.
\end{displaymath}\\
where $\mu(1)=1$. From Eq. (\ref{24}) one notices that the analytic continuation of the prime zeta function can be obtained from the analytic continuation of the Riemann zeta function. Introducing the Jacobi theta function
\begin{equation}
\vartheta(x)=\sum_{n=-\infty}^{\infty}\,e^{-n^{2}\,\pi\,x}, \,\,\,x>0
\label{z2}
\end{equation}
and using the usual definition of the Gamma function, we have
\beq
\Gamma\biggl(\frac{s}{2}\biggr)\,\,\pi^{-\frac{s}{2}}\,\zeta(s)=\frac{1}{s(s-1)}+
\int_{1}^{\infty}dx\,
\vartheta_{1}(x)
\biggl(x^{\frac{s}{2}-1}+x^{-\frac{s}{2}-\frac{1}{2}}\Biggr),
\label{z7}
\eeq
where $\vartheta_{1}(x)=\frac{1}{2}(\vartheta(x)-1)$. The integral that appears in the above equation is convergent for
all values of $s$ and therefore this equation gives the analytic continuation of the Dirichlet series defined by Eq.~(\ref{p2}) to the whole complex $s$-plane. The only singularity is the pole at $s=1$, since the pole at $s=0$ is canceled by the pole of the Gamma function $\Gamma(\frac{s}{2})$. Therefore the analyic extension is valid for $s \in \mathbb{C}\setminus\left\{1\right\}$.

The analytic structure of $P(s)$ for $\Re(s)\leq 1$ is given by the pole and nontrivial zeros of the
Riemann zeta function. First, due to the polar structure of the Riemann zeta function, in equation (\ref{24}) we have the following. For all square free positive integers $k$ on the real axis we get that $s=\frac{1}{k}$ is a singular point. Moreover, one has a clustering of singular points along the imaginary axis from the non-trivial zeros of the Riemann zeta function~\cite{lan}. Hence the prime zeta function has an analytic continuation only in the strip $0<\sigma\leq 1$. Therefore, with the exception of those points, Eq.~(\ref{24}), together with Eq.~(\ref{z7}), gives a representation for $P(s)$ valid for $\Re(s)>0$.

Let us use that $\zeta(ks)=z$, where $z$ is a complex variable. The function $\ln z$ is analytic at
every point of its Riemann surface. We have that $d P(s)/ds = P'(s)$
can be computed for $\Re(s)>0$.
From Eq.~(\ref{24}), one has
\begin{equation}
P'(s)=\sum_{k=1}^{\infty}\mu(k)\frac{\zeta'(ks)}{\zeta(ks)}.
\label{de}
\end{equation}
As mentioned above, the requisite for defining a regularized determinant is to compute the corresponding spectral zeta function and its derivative at $s=0$, in order to obtain a well-defined functional integral associated with the scalar field. In the case of prime numbers spectrum, the functional $W[0]$ is given by
\begin{equation}
W[0]=\frac{1}{2}\ln\,\mu^{2}
P(s)|_{s=0}+\frac{1}{2}\frac{d}{ds}P(s)|_{s=0}.
\end{equation}
Collecting our results we get the following pure non-existence theorem:
\vspace{0,1cm}\\
{\em{The impossibility of extending the definition of the analytic function $P(s)$ to the half-plane $\sigma\leq 0$, implies that free scalar quantum fields with the sequence of prime numbers spectrum cannot exist.}}
\vspace{0,1cm}\\

It follows from the above theorem that, in the weak-coupling regime, the generating functional of the Schwinger functions cannot be defined either, since the generating functional for the free theory is meaningless.
As remarked before, such a result can be extended to higher spin fields, in particular spinor fields, with weak interactions. On the other hand, an intriguing issue is whether this situation persists in non-perturbative regimes.

\section{Discussion and conclusions}

In this Letter we showed that free scalar quantum fields with the sequence of prime numbers spectrum cannot exist. Using the functional integral approach we proved that the sequence of primes numbers is not zeta regularizable. On the other hand, suppose that there is a hypothetical quantum system realized in nature which has an energy spectrum given by the sequence of prime numbers. Hence from the above theorem one is forced to accept that there cannot be a free energy associated with such a system.

Since the one-loop effective energy and the Casimir energy are closely related~\cite{blau},
it is natural to discuss the vacuum energy associated with a theory in which prime numbers appear in the zero point-energy.
Suppose a two-dimensional space-time and a box of size $L$. Assuming that there are suitable boundary conditions such that the vacuum energy is
defined by $\langle 0|H|0\rangle\propto\sum p$ where the sum runs over all prime numbers.
For the usual cases of regularization of ill-defined quantities, it is possible to prove the following statement: if the introduction of a exponential cut-off yields a analytical function with a pole in the origin, the analytic extension of the generalized zeta-function is equivalent to the application of a cut-off with the subtraction of the singular part at the origin~\cite{ss,nami1,nami2}.
Once we accept the advantage of the zeta-function method over the cut-off method, we face a serious problem.
Since the prime zeta function has an analytic continuation only in the strip $0<\sigma\leq 1$, one cannot obtain a finite renormalized vacuum energy for this particular situation. Similar conclusions were obtained in Ref.~\cite{barbero}. One can show that the same situation occurs in an interacting field theory. It is not possible to interpret the generating functional of Schwinger functions nor the one-loop effective action of the model in terms of well-defined mathematical objects.

There is a point that deserves further discussions. We mention that has been suggested in the literature a way to establish the existence of a quantum system possessing the nontrivial zeros of the $\zeta(s)$ function as energy eigenvalues by constructing one-dimensional potentials with a multifractal character. Whether or not this can be extended to fields depends crucially on one's ability to set up such fractal potentials in the quantum field theory framework~\cite{stri,dunne2,dune}.

\section*{Acknowlegements}

We would like to thank J. Stephany, S. Alves Dias and G. Krein for useful discussions. This paper was supported by Conselho Nacional de Desenvolvimento Cient\'ifico e Tecnol{\'o}gico (CNPq).

\end{document}